\documentclass[
 twocolumn,
  prb,
  amsmath,
  amssymb,
  superscriptaddress,
  floatfix
]{revtex4}

\usepackage{bm}
\usepackage{graphicx}

\begin{document}

\title{Interaction-induced criticality in  $\mathbb{Z}_2$ topological insulators}

\author{P.~M.~Ostrovsky}
\affiliation{
 Institut f\"ur Nanotechnologie, Forschungszentrum Karlsruhe,
 76021 Karlsruhe, Germany
}
\affiliation{
 L.~D.~Landau Institute for Theoretical Physics RAS,
 119334 Moscow, Russia
}

\author{I.~V.~Gornyi}
\affiliation{
 Institut f\"ur Nanotechnologie, Forschungszentrum Karlsruhe,
 76021 Karlsruhe, Germany
}
\affiliation{
A.F.~Ioffe Physico-Technical Institute,
 194021 St.~Petersburg, Russia.
}
\author{A.~D.~Mirlin}
\affiliation{
 Institut f\"ur Nanotechnologie, Forschungszentrum Karlsruhe,
 76021 Karlsruhe, Germany
}
\affiliation{
 Inst. f\"ur Theorie der kondensierten Materie,
 Universit\"at Karlsruhe, 76128 Karlsruhe, Germany
}
\affiliation{
Petersburg Nuclear Physics Institute,
 188350 St.~Petersburg, Russia.
}

\date{\today}

\maketitle

\textbf{
Critical phenomena and quantum phase transitions are paradigmatic
concepts in modern condensed matter physics. 
A central example in the field of mesoscopic physics is the
localization-delocalization  (metal-insulator) quantum phase transition
driven by disorder --- the Anderson transition.~\cite{Evers08}
Although the notion of localization has appeared half a century ago,
this field is still
full of surprising new developments. The most recent arenas where novel
peculiar localization phenomena have been studied are 
graphene \cite{CastroNeto09}
and topological insulators,~\cite{Kane05, Kane05a, Bernevig06, Koenig07, Hsieh08}
i.e., bulk insulators with
delocalized (topologically protected) states on their surface.
Besides exciting physical properties, the topological protection
renders such systems promising candidates for a variety of
prospective electronic and spintronic devices.
It is thus of crucial importance to understand properties of boundary metallic modes in the
realistic systems when both disorder and interaction 
are present.
Here we find a novel critical state which emerges in the bulk of
two-dimensional quantum spin Hall (QSH) systems and on the surface of
three-dimensional topological insulators with strong spin-orbit interaction
due to the interplay of nontrivial $\mathbb{Z}_{2}$ topology and the Coulomb repulsion.
At low temperatures, this state possesses a universal value of electrical conductivity.
In particular, we predict that the direct QSH phase transition occurs via this novel state.
Remarkably, the interaction-induced critical state emerges on the surface of a
three-dimensional topological insulator without any adjustable parameters.
This ``self-organized quantum criticality'' is a novel concept in the field of
interacting disordered systems.
}

The
universality of critical phenomena has been studied both in 
the area of strongly correlated systems and in
mesoscopics.
It is now established
that disordered electronic systems can be classified into 10
symmetry classes
(for review see Ref.~\onlinecite{Evers08}).
Very generally, the localization properties are 
determined by the symmetry class and
dimensionality of the system. The critical behavior of a system
depends also on the underlying topology. 
This is particularly relevant for topological insulators.~\cite{Kane05, Kane05a, Bernevig06, Koenig07, Hsieh08, Roth09, Hsieh09, Hsieh09a, Roushan09, Zhang09, Chen09}

The famous example of a topological insulator is a two-dimensional (2D) system on one of quantum Hall
plateaus in the integer quantum Hall effect (QHE). Such a system
is characterized by an integer (Chern number) $n = ...,-2,-1,0,1,2,...$ which
counts the edge states.  Here the sign determines the direction of
chiral edge modes. The integer quantum Hall edge is thus a
topologically protected 1D conductor realizing the group
$\mathbb{Z}$.   

Another ($\mathbb{Z}_2$) class of topological insulators \cite{Kane05, Kane05a, Bernevig06} can
be realized in systems with strong  spin-orbit interaction and
without magnetic field  --- and was discovered in HgTe/HgCdTe structures in
Ref.~\onlinecite{Koenig07} (see also Ref.~\onlinecite{Roth09}).  Such systems were found to possess two distinct insulating
phases, both having a gap in the bulk electron spectrum but
differing by the edge properties.  
While the normal insulating phase has no edge states, the
topologically nontrivial insulator is characterized by a pair of mutually
time-reversed delocalized edge states penetrating the bulk gap.
Such state shows the quantum spin Hall (QSH) effect 
which was theoretically predicted in a model system of
graphene with spin-orbit coupling.~\cite{Kane05a, Sheng05}
The transition between the two topologically nonequivalent phases 
(ordinary and QSH insulators) is driven by inverting the band gap.~\cite{Bernevig06}
The  $\mathbb{Z}_{2}$ topological order is
robust with respect to disorder: since the time-reversal
invariance forbids backscattering of the edge states at the boundary
of QSH insulators, these states are topologically protected from localization.

\begin{figure*}
\centerline{\includegraphics[width=1.4\columnwidth]{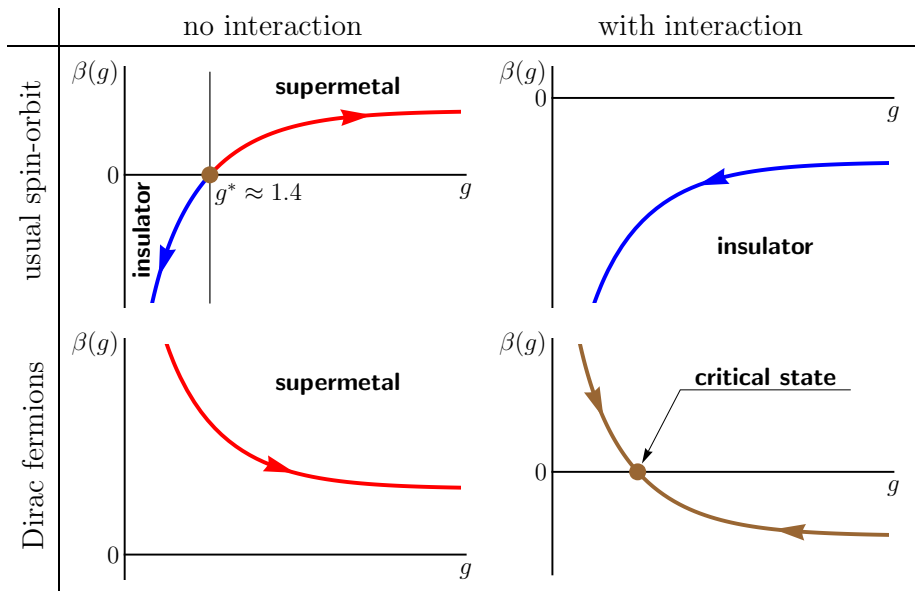}}
\caption{Schematic scaling functions for the conductivity of 2D 
disordered systems of symplectic symmetry class. The plotted beta 
functions $\beta(g)=dg/d\ln L$ determine the flow of the dimensionless 
conductivity $g$ with increasing system size $L$ (as indicated by the 
arrows). The upper two panels show the beta functions for ordinary 
spin-orbit systems
which are not topologically protected (left: no interaction; right: 
Coulomb interaction included).
The lower two panels demonstrate the scaling for topologically 
protected Dirac fermions
(left: no interaction; right: Coulomb interaction included). }
\label{Fig:beta}
\end{figure*}

A related three-dimensional (3D)
$\mathbb{Z}_2$ topological insulator was discovered in Ref. \onlinecite{Hsieh08} where
crystals of Bi$_{1-x}$Sb$_{x}$ were investigated. The boundary in
this case gives rise to a 2D topologically protected metal.
Similarly to 2D topological insulators, the inversion of the 3D band gap induces an odd number 
of the surface 2D modes.~\cite{Fu07} These states in BiSb have been studied
experimentally in Refs.~\onlinecite{Hsieh08, Hsieh09} and \onlinecite{Roushan09}. 
Other examples of 3D topological insulators include BiTe and BiSe systems.~\cite{Zhang09, Chen09, Hsieh09a}

Both in 2D and 3D, $\mathbb{Z}_2$ topological insulators are band insulators with the following
properties: (i) time reversal invariance is preserved (unlike
ordinary quantum Hall systems); 
(ii) there exists a topological
invariant, which is similar
to the Chern number in QHE;
(iii) this invariant belongs to the group $\mathbb{Z}_2$ 
and reflects the presence or absence of delocalized edge modes (Kramers pairs).~\cite{Kane05}
For the sake of completeness we overview the full classification of topological insulators and 
superconductors~\cite{Schnyder08} in Supplementary Information. 

In this paper, we consider the effect of interactions on $\mathbb{Z}_2$
topological insulators belonging to the symplectic symmetry class,
characteristic to systems with strong spin-orbit interaction.
In particular, we predict a novel critical state which emerges 
in a 2D system due to the interplay of nontrivial topology and the Coulomb interaction.

Let us start with reviewing the localization properties of 2D systems of symplectic symmetry class AII
without Coulomb interaction. 
In conventional spin-orbit systems (e.g. semiconductors with spin-orbit scattering),
there are two phases: metal and insulator with the Anderson transition between them,
see Fig.~\ref{Fig:beta}.  A qualitatively different situation occurs in a single species of
massless Dirac fermions in a random scalar potential. This system also belongs to the symplectic symmetry class
but its metallic phase is ``topologically protected''
whatever disorder strength.
This state has been recently predicted for disordered graphene with no spin- and no 
valley-mixing.~\cite{Ostrovsky07,  Ostrovsky07a, Ryu07}
The absence of localization in this model as well as the 
``supermetal'' scaling (Fig.~\ref{Fig:beta}) have been confirmed in numerical
simulations.~\cite{Bardarson07, Nomura07} Although a genuine single Dirac fermion cannot be realized in a 
truly 2D microscopic theory because of the famous ``fermion doubling'' problem, it emerges on the surface
of a 3D topological insulator.~\cite{Fu07}

The 3D topological insulators are characterized by the inverted sign of the gap
(band inversion). This generates the
surface states, as was first pointed out in Ref. \onlinecite{Dyakonov81}. 
The effective 2D surface Hamiltonian has a Rashba form (see the derivation in Supplementary Information)
and describes a single species of 2D massless Dirac particles (cf. Ref.~\onlinecite{Volkov85}). It 
is thus analogous to the Hamiltonian of graphene with just a single valley.
In the absence of interaction, the conductivity of the disordered
surface of a 3D topological insulator therefore scales to infinity with increasing the system
size.
This behavior defines the topologically-protected 
``supermetallic phase'' discussed above.

\begin{figure*}
\centerline{\includegraphics[width=1.7\columnwidth]{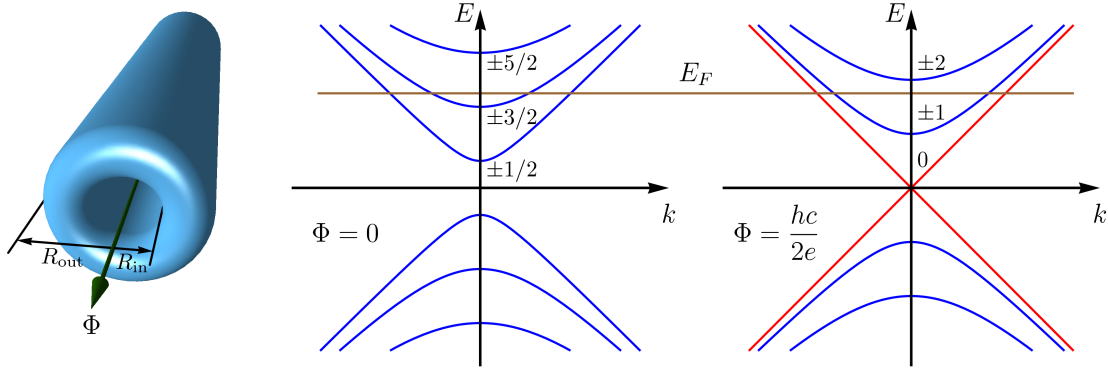}}
\caption{Left: schematic illustration of the hollow cylinder sample used for proving
the absence of localization on the surface of a 3D topological insulator (see Supplementary Information).
Right: the energy spectra of a clean 2D system on the surface of the cylinder 
with zero flux and with half of the magnetic flux quantum
penetrating the cylinder.}
\label{Fig:tbs}
\end{figure*}

Let us now ``turn on'' the Coulomb interaction between electrons. 
Since a topological insulator is
characterized by the presence of non-localizable surface states,
its robustness with respect to interaction
can be reformulated in terms of the absence of
localization of the boundary states. 
At this point it is worth recalling the celebrated example of a 2D
topological insulator, the QHE insulator, in which
the interaction cannot destroy the chiral
1D edge modes on the boundary of a 2D QHE sample.
Furthermore, the two consequent QHE topological insulators (QHE plateaus)
are separated by a delocalized (critical) 2D
state. Since the topological insulator phases are robust, the interaction is not capable of localizing 
electrons in this state separating the two topological insulators.

Having in mind this analogy with the QHE, one can expect
that the delocalized states in $\mathbb{Z}_2$ topological insulators are not destroyed by the
interaction either. Indeed, arguments in favor of the stability of 
$\mathbb{Z}_2$ topological insulators with respect to interactions were given in
Refs.~\onlinecite{Kane05} and \onlinecite{Lee08}.
Below we demonstrate the $\mathbb{Z}_2$ topological order and discuss its 
implications in 2D and 3D interacting systems in the presence of disorder.

We first consider the interacting massless Dirac electrons on the surface of a 3D
topological insulator. Without interaction, the surface states are delocalized in
the presence of arbitrarily strong potential disorder. In Supplementary Information we
demonstrate that the interaction cannot fully localize the surface Dirac fermions.
This is achieved by considering the topological insulator of a hollow cylinder
geometry threaded by half of the magnetic flux quantum, see Fig.~\ref{Fig:tbs}. 
Our 2D problem then reduces
to the quasi-1D model
with an odd number of channels. Full 2D localization would be in contradiction with
known results on the absence of localization in such quasi-1D symplectic wires.
Since delocalization in quasi-1D geometry survives in the presence of interaction,
this is also true for the 2D interacting Dirac electrons on the surface of a 3D
topological insulator.

Can the topologically protected 2D state
be a supermetal ($g \to \infty$) as in the noninteracting case? To answer this question we
employ the perturbative renormalization group (RG) applicable for large 
conductivity $g\gg 1$.

It is well known that 
in a 2D diffusive system the interaction leads to
logarithmic corrections to the conductivity.~\cite{Altshuler85}
These corrections (together with the interference-induced
ones) can be summed up with the use of RG
technique.~\cite{Finkelstein, Punnoose01, Belitz94} The one-loop equation
for renormalization of the conductivity in the symplectic class
with Coulomb interaction has the following form:
\begin{equation}
 \beta(g)=\frac{d g}{d \ln L}=\frac{N}{2}-1+(N^2-1) {\cal F},
\label{RG}
\end{equation}
where $N$ is the number of degenerate species (``flavors'': spin, isospin, \dots) and $L$ is the system size.

The first term, $N/2$, describes the effect of weak antilocalization due to disorder
(this term exists also in the absence of interaction) for $N$
parallel conductors. The second term, $-1$, is induced by the Coulomb
interaction in the singlet channel and has a localizing effect: it
suppresses the conductivity.
The last term on the r.h.s. of Eq.~(\ref{RG}) is due to the
interaction in the multiplet (in the flavor space) channel. This
term yields a positive (antilocalizing) correction to the
conductivity.
The multiplet interaction parameter
${\cal F}$ is itself subject to renormalization.~\cite{Finkelstein}

In the degenerate case $N > 1$ (as, for example, in graphene with no spin-
and valley-mixing where $N =4$), the beta function (\ref{RG}) is positive corresponding to the
``supermetal'' phase. The situation is essentially different for 2D states on the surface of a
3D $\mathbb{Z}_2$ topological insulator where we have a symplectic system with  $N=1$.
According to Eq.~(\ref{RG}), the negative
interaction-induced term in $\beta(g)$ now dominates while the multiplet term is absent. Therefore,
for $g\gg 1$ the conductance decreases upon renormalization.
This means that due to interaction the supermetal fixed point 
becomes repulsive.

Thus, on one hand, we encounter the tendency to
localization due to the interaction. On the other hand, the states
on the surface of the topological insulator are topologically
protected from the localization.
At $g\sim 1$
the topological protection reverses the sign of the $\beta$ function,
similarly to the ordinary QHE. As a result, a critical point
emerges due to the combined effect of interaction and topology, see Fig.~\ref{Fig:beta}.
This type of criticality should be contrasted with the QHE criticality
which exists already without interactions.  
In our case, even in
the absence of the critical state in a non-interacting model, the
criticality is inevitably established in the realistic
interacting systems. This novel interaction-induced critical state is
the major result of our paper.

\begin{figure*}
\centerline{\includegraphics[width=1.8\columnwidth]{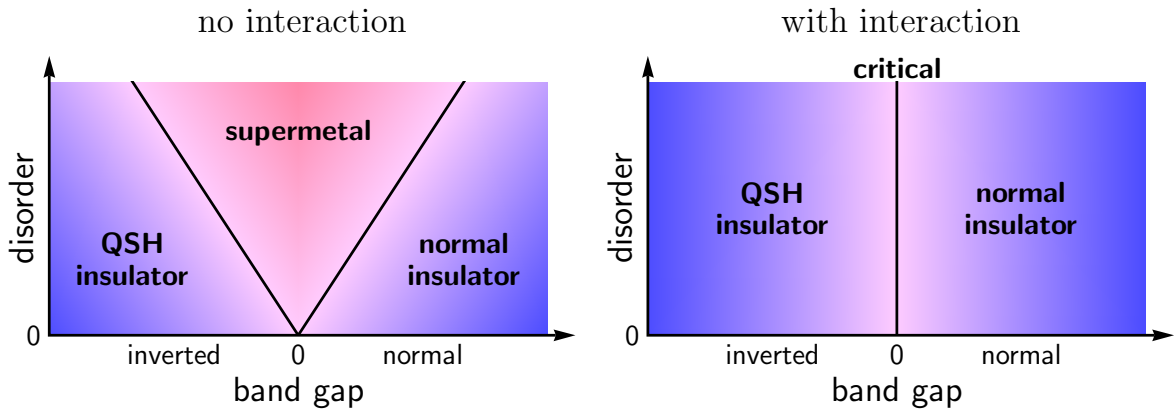}}
\caption{The phase diagrams of a disordered 2D system demonstrating the QSH effect.
Left: noninteracting case. In contrast to the clean case, the two topologically nonequivalent
phases (QSH and ordinary insulators) have been found to be separated by the metallic phase.~\cite{Onoda07,Obuse07}
The transition~\cite{Evers08} between metal and any of the two insulators occurs
at the critical value of dimensionless (in units $e^2/h$) conductivity $g=g^* \approx 1.4$;
both transitions are believed to belong to the same universality class.
For $g<g^*$ all bulk states are eventually localized in the limit of large system, while for $g>g^*$
antilocalization drives the system to the ``supermetallic'' state, $g \to \infty$.
Right: interacting case (with Coulomb interaction not screened by external gates).
Interaction ``kills'' the supermetallic phase. As a result, the two insulating phases are separated by 
the critical line.}
\label{Fig:ph}
\end{figure*}

Let us now return to the case of 2D  $\mathbb{Z}_2$ topological
insulators. As discussed above, without interaction
disorder was found to induce a metallic phase separating the
two (QSH and ordinary) insulating  phases.~\cite{Onoda07,Obuse07} The conductance in
the metallic phase scales to infinity because of the weak
antilocalization specific to the symplectic symmetry class.
The schematic phase diagram for the noninteracting case is shown in Fig.~\ref{Fig:ph} (left panel).

What would change in this phase diagram when interaction is taken
into account? The answer follows from Eq.~(\ref{RG}). The 2D
disordered QSH system contains only a single flavor of particles, $N=1$.
Indeed, the spin-orbit interaction breaks the spin-rotational
symmetry, whereas the valleys are mixed by disorder. 
As a result, the supermetal phase does not survive in the
presence of Coulomb interaction:
at $g\gg 1$ the interaction-induced localization wins.
This is analogous to the case of the surface of a 3D topological
insulators discussed above.

The edge of a 2D topological insulator is protected from the full
localization, as was discussed already in the pioneering works by Kane
and Mele.~\cite{Kane05, Kane05a} In the presence of interaction, the 
counter-propagating edge modes constitute the Luttinger liquid. 
As disorder-induced backscattering in this liquid is forbidden by 
the time-reversal symmetry, the dimensionless conductance of this 1D system 
is equal to unity. This means that the topological distinction between 
the two insulating phases (ordinary and QSH insulator) is not destroyed 
by the interaction, whereas the supermetallic phase separating them disappears. 
Therefore we conclude that the transition between two insulators occurs through 
an interaction-induced critical state, see Fig. ~\ref{Fig:ph} (right panel).

We have thus found that in the two types of systems with strong spin-orbit
coupling the Coulomb interaction induces novel 2D critical states.
This happens, first, at the boundary of 3D topological insulators and, second,
in the bulk of 2D QSH systems, where the critical state separates
the two topologically distinct insulating phases.  In the first case the system can be
described by a 2D interacting symplectic sigma-model with the $\mathbb{Z}_2$ topological
term. The two critical states have much in common:
(i) symplectic symmetry, (ii) $\mathbb{Z}_2$ topological protection,
(iii) interaction-induced criticality, and (iv) conductivity of order unity 
(probably universal). This suggests that the corresponding fixed points might be equivalent.
The existence of the proposed critical states can be verified in
transport experiments on the semiconductor structures with possible gap inversion.

We thank  A. Altland, D. Bagrets, I. Burmistrov, F. Evers, M. Feigel'man, A. Finkel'stein, V. Kagalovsky, D. Khmelnitskii, V. Kravtsov,
I. Lerner, A.W.W. Ludwig, L. Molenkamp, I. Protopopov, and A. Rosch for valuable
discussions. The work was supported by the DFG -- Center for
Functional Nanostructures, by the EUROHORCS/ESF (IVG), and by Rosnauka
grant 02.740.11.5072.

\vspace{1cm}

\section{Supplementary Information}

\textbf{Classification of topological insulators}

\begin{table*}
\caption{Symmetry classes and ``Periodic Table'' of topological insulators.~\cite{Kitaev09,Schnyder08a}
The first column enumerates the symmetry classes of disordered systems which are defined as 
the symmetry classes $H_p$ of the Hamiltonians (second column).
The third column lists the symmetry classes of the classifying spaces (spaces of reduced Hamiltonians).~\cite{Kitaev09}
The fourth column represents the symmetry classes of a compact sector of the sigma-model
manifold. The fifth column displays the zeroth homotopy group $\pi_0(R_p)$ of the classifying space.
The last four columns show the possibility of existence of $\mathbb{Z}$ and $\mathbb{Z}_2$ topological insulators
in each symmetry class in dimensions $d=1,2,3,4$.}
\label{Tab:sym}
\begin{ruledtabular}
\begin{tabular}{clclclclclclclclc}
 \multicolumn{1}{c}{} &&
 \multicolumn{3}{c}{Symmetry classes} &&
\multicolumn{1}{c}{} &&
 \multicolumn{4}{c}{Topological insulators}
\\
 $p$ &&
 $H_p$ & $R_p$ & $S_p$ &&
 $\pi_0(R_p)$ &&
 d=1 & d=2 & d=3 & d=4
\\ \hline \hline
 0
 &&
 AI & BDI & CII
 &&
$\mathbb{Z}$
 &&
 0 & 0 & 0 & $\mathbb{Z}$
\\
 1
&&
 BDI & BD & AII
&&
$\mathbb{Z}_2$
  &&
 $\mathbb{Z}$ & 0 & 0 & 0
\\
 2
&&
 BD & DIII  & DIII
&&
$\mathbb{Z}_2$
  &&
 $\mathbb{Z}_2$ & $\mathbb{Z}$ & 0 & 0
\\
3
 &&
 DIII & AII & BD &&
 0 &&
$\mathbb{Z}_2$ & $\mathbb{Z}_2$ & $\mathbb{Z}$ & 0
\\
 4
&&
 AII & CII & BDI
&&
 $\mathbb{Z}$
&&
0 & $\mathbb{Z}_2$ & $\mathbb{Z}_2$ & $\mathbb{Z}$
\\
 5
&&
 CII & C & AI
&&
 0
&&
$\mathbb{Z}$ & 0 & $\mathbb{Z}_2$ & $\mathbb{Z}_2$
\\
 6
&&
 C & CI & CI
&&
 0
&&
0 & $\mathbb{Z}$ & 0 & $\mathbb{Z}_2$
\\
 7
&&
 CI & AI & C
&&
 0
&&
0 & 0 & $\mathbb{Z}$ & 0
\\
\hline
\hline
$0^\prime$
&&
 A & AIII & AIII
&&
 $\mathbb{Z}$
&&
0 & $\mathbb{Z}$ & 0 &  $\mathbb{Z}$
\\
$1^\prime$
&&
 AIII & A & A
&&
0
&&
$\mathbb{Z}$ & 0 & $\mathbb{Z}$ & 0
\end{tabular}
\end{ruledtabular}

\end{table*}

The full classification (periodic table) of topological insulators 
and superconductors for all ten symmetry classes~\cite{Zirnbauer96, Altland97}
was developed in Refs. \onlinecite{Kitaev09} and \onlinecite{Schnyder08a}.
In this Section we overview the classification of topological insulators  
closely following  Refs. \onlinecite{Kitaev09} and \onlinecite{Schnyder08a} 
and discuss the connection between the classification schemes of these papers. 
Very generally,
the classification of topological insulators in dimension D can be constructed by studying the 
Anderson localization problem in a D-1 disordered system: absence of localization 
of surface states implies the topological character of the insulator.~\cite{Schnyder08a}

All symmetry classes of disordered systems 
\cite{Zirnbauer96, Altland97} can be divided into two groups:
\{A,AIII\} and \{all other\}. The classes of the big group are labeled by
$p=0,1,\dots,7$. Each class is characterized by (i) Hamiltonian
symmetry class $H_p$; (ii) symmetry class of the classifying space used by Kitaev \cite{Kitaev09}  $R_p$; (iii)
symmetry class $S_p$ of the compact sector $\mathcal{M}_F$ of the
sigma-model manifold. The symmetry class $R_p$  of the classifying space of reduced
Hamiltonians characterizes the space of matrices obtained from the
Hamiltonian by keeping all eigenvectors and replacing all positive
eigenvalues by $+1$ and all negative by $-1$.
Note that
\begin{equation}
 R_p=H_{p+1}, \quad S_p=R_{4-p}.
\label{RHS}
\end{equation}
Here and below cyclic definition of indices $\{0,1,\dots,7\}$ (mod
8) and $\{0',1'\}$ (mod 2) is assumed.

For the classification of topological insulators it is important
to know homotopy groups $\pi_d$ for all symmetry classes. In Table \ref{Tab:sym}
we list $\pi_0(R_p)$; other $\pi_d$ are given by
\begin{equation}
 \pi_d(R_p)=\pi_0(R_{p+d}).
\label{pidRp}
\end{equation}
The homotopy groups $\pi_d$ have periodicity 8 (Bott periodicity).

There are two ways to detect topological insulators:
by inspecting the topology of (i) classifying space $R_p$
or of (ii) the sigma-model space $S_p$.

(i) Existence of topological insulator (TI) of class $p$ in $d$ dimensions
is established by the homotopy group $\pi_0$ for the classifying space  $R_{p-d}$:
\begin{equation}
\begin{cases}
       \text{TI of the type}\ \mathbb{Z}   \\
        \text{TI of the type}\ \mathbb{Z}_2  \end{cases}
\Longleftrightarrow \quad
\pi_0(R_{p-d}) = \begin{cases}
        \mathbb{Z} &  \\
          \mathbb{Z}_2 &
                  \end{cases}
\end{equation}

(ii) Alternatively, the existence of topological insulator of
symmetry class $p$ in $d$ dimensions can be inferred from the
homotopy groups of the sigma-model manifolds, as follows:
\begin{equation}
\begin{cases}
       \text{TI of the type}\ \mathbb{Z} \ \Longleftrightarrow  &\pi_d(S_{p}) = \mathbb{Z}\\
        \text{TI of the type}\ \mathbb{Z}_2  \Longleftrightarrow  &\pi_{d-1}(S_{p}) =\mathbb{Z}_2
\end{cases}
\end{equation}
This criterium is obtained if one requires existence of
``non-localizable'' boundary excitations. This may be guaranteed
by either Wess-Zumino term in $d-1$ dimensions [which is
equivalent to the $\mathbb{Z}$ topological term in $d$ dimensions,
i.e. $\pi_d(S_{p}) = \mathbb{Z}$] for a QHE-type topological
insulator, or by the $\mathbb{Z}_2$ topological term in $d-1$
dimensions  [i.e. $\pi_{d-1}(S_{p}) = \mathbb{Z}_2$] for a QSH-type
topological insulator.

The above criteria (i) and (ii) are equivalent, since
\begin{equation}
 \pi_d(S_p)=\pi_d(R_{4-p})=\pi_0(R_{4-p+d}).
\label{pidSp}
\end{equation}
and
\begin{equation}
 \pi_0(R_p)=  \begin{cases}
        \mathbb{Z} &  \text{for}\ p=0,4,\\
          \mathbb{Z}_2 & \text{for}\ p=1,2.
                  \end{cases}
\label{pi0Rp}
\end{equation}

In this paper we focus on 2D systems of symplectic (AII) symmetry 
class. One sees that this is the only symmetry class out of ten classes that supports 
the existence of $\mathbb{Z}_2$ topological insulators both in 2D and 3D.
The effect of interaction on $\mathbb{Z}_2$ topological
insulators and superconductors in classes DIII (in 2D) and CII (in 3D) will be considered elsewhere.

\vspace{0.5cm}
\textbf{Surface states of 3D topological insulators}

In this section we consider the surface of 3D topological insulators and
the derive the effective surface Hamiltonian.
The realistic microscopic Hamiltonian of such systems
with strong spin-orbit interaction can be modelled 
by the 3D massive Dirac Hamiltonian, 
see, e.g., Refs.~\onlinecite{Fu07} and \onlinecite{Schnyder08}.
We start with the general form of a 3D Dirac Hamiltonian
      \begin{equation}
        H_{3D}= \begin{pmatrix}
                -M & {\mathbf \sigma} {\mathbf p} \\
                 {\mathbf \sigma} {\mathbf p} & M
              \end{pmatrix}.
              \label{3Dhamiltonian}
      \end{equation}
This Hamiltonian is a matrix in the $4\times 4$ space formed by
the spin and pseudospin (sublattice) spaces. 
The interface between the semiconductor and the vacuum is
described by sending the mass $M$ to infinity.

Let us consider a flat interface at
$x=0$. The edge state with zero energy decays into the bulk ($x<0$):
  \begin{equation}
  \Psi=e^{Mx} \begin{pmatrix}
            \psi\\
            \chi
          \end{pmatrix}.
  \end{equation}
Acting  by the Hamiltonian
(\ref{3Dhamiltonian}) on this wavefunction, we obtain the relation between the two
components of the spinor, 
$\chi=i \sigma_x \psi.$
For an arbitrary surface characterized by the normal vector ${\mathbf n}$, 
the general boundary condition reads
\begin{equation}
 \chi=i {\mathbf \sigma} {\mathbf n} \psi.
\label{BC}
\end{equation}
From Eqs.~(\ref{3Dhamiltonian}) and (\ref{BC}) we obtain the
effective 2D surface Hamiltonian in the form of the Rashba
Hamiltonian in a curved space 
  \begin{equation}
  H_{\rm surf} =
  \frac{i}{2}\ [{\mathbf \sigma} {\mathbf p},{\mathbf \sigma} {\mathbf n}]
  = \frac{\nabla {\mathbf n}}{2}
  +\frac{1}{2}\
  \left( {\mathbf n} [{\mathbf p} \times {\mathbf \sigma}]
  + [{\mathbf p} \times {\mathbf \sigma}]{\mathbf n}\right).
  \label{2Dhamiltonian}
    \end{equation}
This 2D Hamiltonian describes a single species of 2D massless Dirac particles and 
is thus analogous to the Hamiltonian of graphene with just a single valley.

\vspace{0.5cm}
\textbf{Absence of localizations of 2D surface states}

Consider the 2D system formed at the surface of a 3D topological insulator and described by the
Hamiltonian (\ref{2Dhamiltonian}). Let us first prove the absence of
localization in the non-interacting case. Assume that all the states in 2D are
localized with some localization length $\xi$. Consider a hollow cylinder with
all dimensions much larger than $\xi$ pierced by the Aharonov-Bohm magnetic flux
$\Phi = hc/2e$ (half of the flux quantum). This value of $\Phi$ does not break
the time-reversal symmetry leaving the system in the symplectic class. In the
absence of disorder we can characterize the surface states by the momentum $k$
along the cylinder axis and by the integer angular momentum $n$. The energy of
such a state is given by
\begin{equation}
 E = \sqrt{k^2 + (n/R)^2}.
\end{equation}
The channels with positive and negative $n$ are degenerate, while $n = 0$
channel is not. Thus the cylinder sustains an odd number of conducting channels
both on the inner and outer surface at any value of chemical potential.

Let us now include disorder and show the absence of localization in
quasi-one-dimensional (q1D) symplectic system with odd number of channels.
The scattering matrix of such a q1D wire has the form
\begin{equation}
 S
  = \begin{pmatrix}
     r & t' \\
     t & r'
    \end{pmatrix}
\end{equation}
with transmission and reflection amplitudes as its entries. The blocks $r$ and
$r'$ are square matrices of the size determined by the number of channels.
Time-reversal symmetry of the symplectic type imposes the following restrictions
on the amplitudes entering the matrix $S$:
\begin{equation}
 r = -r^T, \quad r' = -{r'}^T, \quad t = {t'}^T.
\end{equation}
Calculating the determinant of the both sides of the first identity and taking
into account the odd size of the matrix $r$, we obtain $\det r = 0$. This
implies a zero eigenvalue of $r$ and hence the existence of a channel with
perfect transmission. We conclude: in a q1D wire of symplectic symmetry with an
odd number of channels one channel always remains delocalized.~\cite{Zirnbauer92, Mirlin94, Ando, Takane}

Applying the q1D result to the cylinder constructed above we immediately come
to the controversy: in spite of assumed 2D localization on the surface, the
infinitely long cylinder possesses two (inner and outer) conducting channels.
This proves the absence of localization in 2D.

The proof can be generalized to include the Coulomb interaction. We assume the
temperature to be much smaller than 
the inverse time of electron propagation
through the system. At such low temperatures the inelastic scattering of
electrons is negligible and we can describe the transport by the scattering
matrix calculated at the Fermi energy and accounting for virtual processes.
The symmetry properties of this $S$ matrix are unchanged and hence the above
proof applies.


\vspace*{-0.3cm}

\begin{thebibliography}{30}

\bibitem{Evers08}   F.\ Evers and A.\ D.\ Mirlin, Rev. Mod. Phys. \textbf{80}, 1355 (2008).


\bibitem{CastroNeto09}  A.\ H.\ Castro Neto, F.\ Guinea, N.\ M.\ R.\ Peres,
K.\ S.\ Novoselov, and A.\ K.\ Geim, Rev. Mod. Phys.  {\bf 81},
109 (2009).

\bibitem{Kane05} C.\ L.\ Kane and E.\ J.\ Mele, Phys. Rev. Lett. \textbf{95}, 146802 (2005).

\bibitem{Kane05a} C.\ L.\ Kane and E.\ J.\ Mele, Phys. Rev. Lett. \textbf{95}, 226801 (2005).

\bibitem{Bernevig06} B.\ A.\ Bernevig, T.\ L.\ Hughes, and S.-C.\ Zhang,
Science \textbf{314}, 1757 (2006).

\bibitem{Koenig07} M.\ K\"onig, S.\ Wiedmann, C.\
Br\"une, A.\ Roth, H.\ Buhmann, L.\ W.\ Molenkamp, X.-L.\ Qi, and
S.-C.\ Zhang, Science \textbf{318}, 766 (2007).

\bibitem{Hsieh08} D.\ Hsieh, D.\ Qian, L.\ Wray, Y.\ Xia, Y.\ S.\ Hor, R.\ J.\ Cava, and M.\ Z.\ Hasan,
Nature \textbf{452}, 970 (2008).

\bibitem{Roth09}
A.\ Roth, C.\ Br\"une, H.\ Buhmann, L.\ W.\ Molenkamp, J.\ Maciejko, X.-L.\ Qi, and S.-C.\ Zhang,
Science \textbf{325}, 294 (2009).

\bibitem{Hsieh09} D.\ Hsieh, Y.\ Xia, L.\ Wray, D.\ Qian, A.\ Pal , J.\ H.\ Dil, J.\ Osterwalder,
F.\ Meier, G.\ Bihlmayer, C.\ L.\ Kane, Y.\ S.\ Hor, R.\ J.\ Cava, and M.\ Z.\ Hasan,
Science \textbf{323}, 919 (2008).

\bibitem{Roushan09}
P. Roushan, J. Seo, C. V. Parker, Y. S. Hor, D. Hsieh, D. Qian, A. Richardella, M. Z. Hasan, R. J. Cava, and
A. Yazdani, Nature \textbf{460}, 1106 (2009).

\bibitem{Hsieh09a} D. Hsieh, Y. Xia, D. Qian, L. Wray, J. H. Dil, F. Meier, J. Osterwalder, L. Patthey, J. G. Checkelsky, N. P. Ong, 
A. V. Fedorov, H. Lin, A. Bansil, D. Grauer, Y. S. Hor, R. J. Cava, and  M. Z. Hasan, 
Nature \textbf{460}, 1101 (2009).

\bibitem{Zhang09} H. Zhang, C.-X. Liu, X.-L. Qi, X. Dai, Z. Fang, and S.-C. Zhang,
Nature Phys. \textbf{5}, 438 (2009) 

\bibitem{Chen09} Y.\ L.\ Chen, J.\ G.\ Analytis, J.-H.\ Chu, Z.\ K.\ Liu, S.\ K.\ Mo, X.\ L.\ Qi, H.\ J.\ Zhang, 
D.\ H.\ Lu, X.\ Dai, Z.\ Fang, S.\ C.\ Zhang, I.\ R.\ Fisher, Z.\ Hussain, and Z.-X.\ Shen, Science \textbf{325}, 178 (2009).


\bibitem{Sheng05}
L.\ Sheng, D.\ N.\ Sheng, C.\ S.\ Ting, and F.\ D.\ M.\ Haldane,
Phys. Rev. Lett. \textbf{95}, 136602 (2005).

\bibitem{Fu07} L.\ Fu and C.\ L.\ Kane, Phys. Rev. B \textbf{76}, 045302 (2007).


\bibitem{Schnyder08} A.\ P.\ Schnyder, S.\ Ryu, A.\ Furusaki, and A.\ W.\ W.\ Ludwig,
Phys. Rev. B \textbf{78}, 195125 (2008).

\bibitem{Ostrovsky07}  P.\ M.\ Ostrovsky, I.\ V.\ Gornyi, and A.\ D.\ Mirlin,
  Phys. Rev. Lett. \textbf{98}, 256801 (2007).

\bibitem{Ostrovsky07a}  P.\ M.\ Ostrovsky, I.\ V.\ Gornyi, and A.\ D.\ Mirlin,
Eur. Phys. J. Special
Topics \textbf{148}, 63 (2007).

\bibitem{Ryu07}
S.\ Ryu, C.\ Mudry, H.\ Obuse, and A.\ Furusaki, Phys. Rev. Lett. \textbf{99}, 116601 (2007).

\bibitem{Bardarson07} J.\ H.\ Bardarson, J.\ Tworzydlo, P.\ W.\ Brouwer, and
  C.\ W.\ J.\ Beenakker, Phys. Rev. Lett. \textbf{99}, 106801
    (2007).

\bibitem{Nomura07} K.\ Nomura, M.\ Koshino, and S.\ Ryu, Phys. Rev. Lett. \textbf{99},
    146806 (2007).


\bibitem{Dyakonov81} M.\ I.\ Dyakonov and A.\ V.\ Khaetskii, 
JETP Lett. \textbf{33}, 115 (1981).

\bibitem{Volkov85}
B.\ A.\ Volkov and O.\ A.\ Pankratov,  JETP Lett. \textbf{42}, 178 (1985).

\bibitem{Lee08}
S.-S.\ Lee and S.\ Ryu, Phys. Rev. Lett. \textbf{100}, 186807 (2008).



\bibitem{Altshuler85} B.L.~Altshuler and A.G.~Aronov, in {\em
    Electron-electron interactions
in disordered conductors}, edited by A.L.~Efros and M.~Pollak
    (Elsevier, 1985), p. 1.

\bibitem{Finkelstein} A.\ M.\ Finkelstein, 
Sov. Sci. Rev. A. Phys. {\bf 14}, 1 (1990).

\bibitem{Punnoose01} A.\ Punnoose and  A.\ M.\ Finkel'stein,
Phys. Rev. Lett. \textbf{88}, 016802 (2001).


\bibitem{Belitz94} D.\ Belitz and T.\ R.\ Kirkpatrick, Rev. Mod. Phys. \textbf{66}, 261 (1994).


\bibitem{Onoda07}
M.\ Onoda, Y.\ Avishai, and N.\ Nagaosa, Phys. Rev. Lett.
\textbf{98}, 076802 (2007).

\bibitem{Obuse07}  H.\ Obuse, A.\ Furusaki, S.\ Ryu, and C.\ Mudry, Phys. Rev. B \textbf{76},
  075301 (2007).

\bibitem{Zirnbauer96} M.\ R.\ Zirnbauer, J. Math. Phys. \textbf{37}, 4986 (1996).

\bibitem{Altland97}
A.\ Altland and M.\ R.\ Zirnbauer, Phys. Rev. B \textbf{55}, 1142 (1997).

\bibitem{Schnyder08a} A.\ P.\ Schnyder, S.\ Ryu, A.\ Furusaki, and A.\ W.\ W.\ Ludwig,
AIP Conf. Proc. \textbf{1134}, 10 (2009).

\bibitem{Kitaev09} A.\ Yu.\ Kitaev,  AIP Conf. Proc. \textbf{1134}, 22 (2009).


\bibitem{Zirnbauer92}
 M.R.~Zirnbauer, Phys. Rev. Lett. {\bf 69}, 1584-1587 (1992).

\bibitem{Mirlin94}
A.D.~Mirlin, A. M\"uller-Groeling, and M.R. Zirnbauer, Ann. Phys. {\bf 236},
325 (1994).

\bibitem{Ando}
T.~Ando and H.~Suzuura, J. Phys. Soc. Jpn. {\bf 71}, 2753 (2002).

\bibitem{Takane}
 Y.~Takane, J. Phys. Soc. Jpn. {\bf 73}, 1430 (2004). 


\end{thebibliography}
\end{document}